\documentclass[11pt]{article}

\usepackage{amsmath,amssymb,amsthm}
\usepackage[margin=2.5cm]{geometry}
\usepackage{hyperref}
\usepackage{color}

\newcounter{Theorems}
\newcounter{Definitions}
\newcounter{Conjectures}
\begin{document}
\begin{titlepage}
\begin{flushright}

\end{flushright}

\begin{center}
{\Large\bf $ $ \\ $ $ \\
%TITLE
Insertion of vertex operators using BV formalism
}\\
\bigskip\bigskip\bigskip
{\large Andrei Mikhailov}
\\
\bigskip\bigskip
{\it Instituto de Fisica Teorica, Universidade Estadual Paulista\\
R. Dr. Bento Teobaldo Ferraz 271, 
Bloco II -- Barra Funda\\
CEP:01140-070 -- Sao Paulo, Brasil\\
}

\vskip 1cm
\end{center}

\begin{abstract}
%ABSTRACT
We develop a general framework for the insertion of
                 vertex operator on the string worldsheet, in BV formalism. Such
                 insertions correspond to deformations of the Master Action which
                 breaks the gauge symmetry to a subgroup, and then restoring the full
                 gauge symmetry by integrating over a cycle in the space of Lagrangian
                 submanifolds.  We provide the general construction, global on the moduli space, which was previously conjectured
                 in a form local on the worldsheet.  We explain how the enhancement of
                 the gauge symmetry in equivariant BV formalism can be seen as an application
                 of the general idea of BV effective action.
                 We derive an integral formula for the deformation
                 of the contraction operator due to the vertex insertion.
\end{abstract}

\vfill
{\renewcommand{\arraystretch}{0.8}%
}

\end{titlepage}

\tableofcontents

\section{Introduction}\label{Introduction}

String amplitudes are computed in perturbation theory by integrating the CFT correlators
of vertex operators over the moduli space of Riemann surfaces.
A mathematical abstraction for the string measure within the BV formalism was outlined in
\cite{Schwarz:2000ct}
and further developed in \cite{Mikhailov:2016myt},\cite{Mikhailov:2016rkp}.
An important role is played by the equivariant BV formalism developed in
\cite{Nersessian:1993me}, \cite{Nersessian:1993eq}, \cite{Nersessian:1995yt},
\cite{Getzler:2015jrr}, \cite{Getzler:2016fek}, \cite{Cattaneo:2016zrn}, \cite{Getzler:2018sbh},
\cite{Bonechi:2019dqk}. In particular, the relation to some differential graded Lie superalgebras
introduced in \cite{Alekseev:2010gr} was discussed in \cite{Mikhailov:2020lzo}. 

In this paper we will continue the study of equivariant BV in the context of string measure.
We will mostly concentrate on the ``covariant insertion'' of vertex operators
on the string worldsheet \cite{Polchinski:1988jq},\cite{Nelson:1988ic}.
Our approach emphasizes the gauge symmetries of the string sigma-model, which include worldsheet diffeomorphisms.
In BV formalism, gauge symmetries are exact in the following sense. If $\xi$ is an infinitesimal
gauge symmetry, then its BV Hamiltonian $\underline{\xi}$ is:
\begin{equation}
   \underline{\xi} = \Delta \underline{i_1\langle \xi\rangle}
   \label{IntroExactCg}\end{equation}
where $\Delta$ is the BV Odd Laplace operator
(roughly speaking it is the odd Poisson bracket with the BV Master Action, $\{S_{\rm BV},\_\}$)
and $\underline{i_1\langle\xi\rangle}$ some function on the BV phase space linearly dependent on $\xi$ as a parameter.
For example, in the case of bosonic string, $\xi$ is a vector field on the string worldsheet $\Sigma$,
and $i_1$ is
\begin{equation}
     \underline{i_1\langle\xi\rangle} = \int_{\Sigma} c^{\star}_{\alpha}\xi^{\alpha}
     \label{IntroI1}\end{equation}
where $c^{\star}$ is the antifield of the $c$-ghost. Physical states of the bosonic string
correspond to the ghost number two vertex operators. For example, the tachyon vertex is:
\begin{equation}
   V = c\bar{c} e^{ikx}
   \end{equation}
This is inserted at a point on the string worldsheet, but one should also integrate over the position
of the insertion. The integration procedure involves ``stripping'' the $c$-ghosts:
\begin{equation}
   \int_{\Sigma} d^2z e^{ikx} = \int_{\Sigma} U
   \end{equation}
where the \emph{integrated} vertex operator (a two-form on $\Sigma$) is defined as follows:
\begin{equation}
     U = \{c^{\star}dz,\{\bar{c}^{\star} d\bar{z},V\}\}
     \label{IntroIntegratedVertex}\end{equation}
When we insert the ``unintegrated'' vertex operator $V$, the diffeomorphism symmetry gets broken down
to those maps which preserve the point of insertion.
Then, when we integrate over the worldsheet metrics, we effectively integrate over the position
of the insertion point, and the full diffeomorphism symmetry is restored.

The relation between Eq. (\ref{IntroIntegratedVertex}) and Eq. (\ref{IntroI1}) can be generalized to other
string theories where Eq. (\ref{IntroI1}) is more complicated:
\begin{align} \underline{\xi}\;=\;
 & \Delta \underline{i(\xi)} + {1\over 2}\underline{[i(\xi),i(\xi)]}
\label{IntroExactDg} \\ i(\xi) \;=\;
 & i_1\langle\xi\rangle + i_2\langle\xi\otimes\xi\rangle + \ldots
\nonumber{} \end{align}
(We use angular brackets around the argument of a function, when the function is linear.)
This was previously discussed in \cite{Mikhailov:2020lzo}, but the consideration was local on the
space of metrics, and the analogue of Eq. (\ref{IntroIntegratedVertex}) was only conjectured.
Here we derive the ``global'' relation combining the ``vertical'' integration
over the position of the insertion point with the integration over the moduli space of Riemann surfaces.

We do not know explicit examples of string theories where $i_{\geq 2}$ is nonzero, but we suspect that
this is what happens in the pure spinor formalism
(see Section \ref{DeformationOfI}).
One example with nonzero $i_2$ was recently discussed
(not for a string sigma-model) in \cite{Bonechi:2022aji}.

\section{Half-densities and their deformations}\label{HalfDensitiesAndDeformations}

\subsection{Odd Laplace operator}\label{sec:OddLaplaceOperator}

Let $M$ be a supermanifold with an odd symplectic structure.
We call it the ``BV phase space''.
Let $\mathfrak a$ be a Lie superalgebra of functions on $M$, with
opposite statistics, the Lie operation being the BV Poisson bracket $\{,\}$.

In Darboux coordinates:
\begin{equation}
   \{\phi^{\star}_A,\phi^B\} = \delta_A^B
   \label{DarbouxCoordinates}\end{equation}
Half-density $\rho_{1/2}$ on $M$ is a scalar function transforming under diffeomorphisms of $M$
as a square-root of volume element. There is a canonical second order differential operator
$\Delta_{\rm can}$ acting on densities \cite{Khudaverdian:1999}.

A half-density $\rho_{1/2}$ is said to satisfy the Quantum Master Equation (QME) if:
\begin{equation}
   \Delta_{\rm can} \rho_{1/2} = 0
   \end{equation}
Let us assume that $\rho_{1/2}$ is everywhere non-zero.
Then a small deformation of $\rho_{1/2}$ can be described by a function $f$ on $M$:
\begin{equation}
   \delta \rho_{1/2} = f\rho_{1/2}
   \end{equation}
If $\rho_{1/2}$ satisfies QME, the deformation satisfies QME iff:
\begin{align}  
 & \Delta_{\rho_{1/2}} f = 0
\nonumber{} \end{align}
where:
\begin{equation}
   \Delta_{\rho_{1/2}} f = {1\over \rho_{1/2}}\Delta_{\rm can}(f\rho_{1/2})
   \label{DeltaFromRho}\end{equation}
The operator $\Delta_{\rho_{1/2}}$ acts on functions. It does depend on $\rho_{1/2}$,
but its \emph{leading symbol} does not depend on $\rho_{1/2}$. In other words, if $\rho_{1/2}^{(1)}$
and $\rho_{1/2}^{(2)}$ are two half-densities, then $\Delta_{\rho_{1/2}^{(1)}} - \Delta_{\rho_{1/2}^{(2)}}$
is a first order differential operator.

In Darboux coordinates (see Eq. (\ref{DarbouxCoordinates})):
\begin{equation}
   \Delta_{\rho_{1/2}} = \sum_A (-1)^{\bar{A} + 1}{\partial\over\partial\phi_A^{\star}}{\partial\over\partial\phi^A} + \ldots
   \end{equation}
In physics notations:
\begin{align} \Delta_{\rho_{1/2}} \;=\;
 & \Delta^{(0)} + \{S_{\rm BV},\_\}
\nonumber{} \\ \mbox{where\hspace{2.00000ex}}
 & \Delta^{(0)} = \sum_A (-1)^{\bar{A} + 1}{\partial\over\partial\phi_A^{\star}}{\partial\over\partial\phi^A}
\nonumber{} \end{align}

\subsection{Deformation complex}\label{sec:DeformationComplex}

The cohomology of $\Delta_{\rho_{1/2}}$ corresponds to \emph{deformations} of $\rho_{1/2}$ preserving the QME, modulo
trivial deformations.
(Trivial deformations are those which can be undone by an infinitesimal canonical transformations.)

Eq. (\ref{DeltaFromRho}) associates to every half-density $\rho_{1/2}$ a nilpotent second order
differential operator with the leading symbol $\Delta^{(0)}$, which is determined by the odd Poisson structure.
(The leading symbol \emph{is} the odd Poisson structure.)
In fact, this correspondence is one-to-one up to the multiplication of $\rho_{1/2}$ by a constant.
For each nilpotent second order differential operator with leading symbol $\Delta^{(0)}$ exists a half-density
$\rho_{1/2}$, satisfying the QME, such that this operator is $\Delta_{\rho_{1/2}}$.

Therefore there is one-to-one correspondence between deformation complexes and half-densities.
The differential of the deformation complex
is $d = \Delta^{(0)} + \{S_{\rm BV},\_\}$, and the half-density is $e^{S_{\rm BV}}$.
The correspondence between deformation complex and half-density
may be called ``exponentiation of the BV Hamiltonian of differential'':
\begin{equation}
   \Delta^{(0)} + d \mapsto e^{\underline{d}}
   \label{ExpD}\end{equation}
where $\underline{d}$ denotes the BV Hamiltonian of the vector field $d$
(which is assumed to preserve the odd symplectic form).
We must stress that Eq. (\ref{ExpD}) only holds in Darboux coordinates.
The operator $\Delta^{(0)} + d$ is a second order differential operator, not a vector field.
And $e^{\underline{d}}$ is actually a half-density, not a function.
We will use this ``physics notations'' throughout this paper, remembering that this $e^{S_{\rm BV}}$ is actually a half-density.

We will need a slight generalization of the correspondence (\ref{ExpD}).
Let $M$ be a BV phase space, $\mbox{Fun}(M)$ denote smooth functions on $M$, and $W$ some differential
commutative superalgebra.
Suppose that $F\in \mbox{Fun}(M)\otimes W$ satisfies:
\begin{equation}
   \left(\Delta^{(0)} + \{F,\_\} + d_W\right)^2 = 0
   \end{equation}
Then there is a half-density $e^F\in \mbox{Dens}_{1/2}\otimes W$ (with the appropriate completion of the tensor product)
satisfying:
\begin{equation}
   (\Delta_{\rm  can} + d_W)e^F = 0
   \end{equation}
Half-densities (just as functions) can be pulled-back by diffeomorphisms.
In case of a function $f$ and a diffeomorphism $g$, the pull-back is a composition, therefore it is natural
to denote it $f\circ g$. We will use the same notation for the pull-back of a half-density:
\begin{equation}
   e^{S_{\rm BV}}\circ g
   \end{equation}
If $e^{S_{\rm BV}}$ satisfies QME, and $g$ is a canonical transformation deformable to the identity transformation, then:
\begin{equation}
   e^{S_{\rm BV}}\circ g = e^{S_{\rm BV}} + \Delta_{\rm can}(\ldots)
   \end{equation}

\section{DGLA $Dh$  and equivariant BV}\label{EquivariantBV}

\subsection{Cone and Weil algebra}\label{sec:ConeAndWeil}

Let $\mathfrak h$ be a Lie superalgebra, $C{\mathfrak h}$ be the cone of $\mathfrak h$, and $W_{\mathfrak h}$ be the Weil algebra of $\mathfrak{h}$.
(See \textit{e.g.} \cite{Meinrenken} for the definitions and general introduction.)
The cone $C\mathfrak h$ is generated by ${\cal L}\langle x \rangle$ and $\iota\langle x\rangle$ where $x\in \mathfrak{h}$. We use angular brackets $\langle\ldots\rangle$ instead of the usual $(\ldots)$ to emphasize linear dependence of a function on its argument. 
The generators of $W_{\mathfrak{h}}$ will be denoted ${\bf\theta}^a$ and ${\bf t}^a$, and its differential will
be denoted $d_W$:
\begin{align} d_W{\bf \theta}^a =
 & {\bf t}^a - {1\over 2} f^a_{bc} {\bf \theta}^b{\bf \theta}^c
\nonumber{} \\ d_W{\bf t}^a =
 & f^a_{bc}{\bf\theta}^b{\bf t}^c
\nonumber{} \end{align}
The action of $C\mathfrak{h}$ on $W_{\mathfrak{h}}$ is such that $\iota\langle x\rangle$
acts as $x^a{\partial\over\partial{\bf \theta}^a}$.

\subsection{Definition of $D\mathfrak h$}\label{sec:DefDg}

We use the DGLA  $D\mathfrak h$ as it was defined in \cite{Alekseev:2010gr},
and its straightforward generalization to Lie superalgebras, \cite{Mikhailov:2020lzo}.
It is generated by $l_a$ and $i_{a_1\ldots a_n}$ where $n\in\{1,2,\ldots\}$.
The $l_a$ generate $\mathfrak{h}$.
The $i_{a_1\ldots a_n}$ are symmetric in $a_1,\ldots,a_n$;
they generate a free Lie superalgebra.
We denote:
\begin{equation}
   i_n\langle x\otimes\cdots\otimes x\rangle = i_{a_1\ldots a_n} x^{a_1}\cdots x^{a_n}\;,\;
   i(x) = \sum_n i\langle x^{\otimes n}\rangle
   \end{equation}
The differential acts as follows:
\begin{align}  
 & d_{D\mathfrak{h}} l\langle x\rangle = 0
\nonumber{} \\  
 & d_{D\mathfrak{h}} i(x) = l\langle x\rangle + {1\over 2}[i(x),i(x)]
\nonumber{} \end{align}
The characteristic property of the construction $\mathfrak{h} \mapsto D\mathfrak{h}$ is:
\begin{itemize}\item For any linear space $V$,
            to define an action of $C{\mathfrak h}$ on $W_{\mathfrak h}\otimes V$ 
            is the same as to define the action of $D{\mathfrak h}$ on $V$.\end{itemize}
More precisely, we must assume that $\iota\langle x\rangle$ acts on $W_{\mathfrak{h}}\otimes V$
as $x^a{\partial\over\partial{\bf\theta}^a}$ and ${\cal L}\langle x\rangle$ acts as
$f^a_{bc}x^b\left({\bf\theta}^c{\partial\over\partial{\bf\theta}^a} + {\bf t}^c{\partial\over\partial{\bf t}^a}\right) + x^ar_V\langle l_a\rangle$ where $r_V\langle l_a\rangle$ are some action of $\mathfrak{h}$ on $V$.
The most nontrivial part of the construction is to define the action of $d_{C\mathfrak{h}}$ on $W_{\mathfrak{h}}\otimes V$,
which would be compatible with $[\iota\langle x\rangle, d_{C\mathfrak{h}}] = {\cal L}\langle x\rangle$. This is where $i(x)$ enters, in the last term of Eq. (\ref{EntersI}):
\begin{align} r_{W_{\mathfrak h}\otimes V}\langle \iota\langle x\rangle\rangle\;=\;
 & x^a{\partial\over\partial{\bf\theta}^a}
\nonumber{} \\ r_{W_{\mathfrak h}\otimes V}\langle {\cal L}\langle x\rangle\rangle\;=\;
 & f^a_{bc}x^b\left({\bf\theta}^c{\partial\over\partial{\bf\theta}^a} + {\bf t}^c{\partial\over\partial{\bf t}^a}\right) + x^ar_V\langle l_a\rangle
\nonumber{} \\ r_{W_{\mathfrak h}\otimes V}\langle d_{C\mathfrak{h}}\rangle\;=\;
 & d_{W\mathfrak{h}} + r_V\langle d_{D\mathfrak{h}}\rangle + r_V\langle l\langle{\bf \theta}\rangle\rangle
             - r_V\langle i({\bf t})\rangle
\label{EntersI} \end{align}

\subsection{Action of $D\mathfrak{h}$ on BV phase space}\label{sec:DgOnBV} 

Let us take $V = {\mathfrak a}$, where $\mathfrak{a}$ is the Lie superalgebra of functions on BV phase space $M$,
and require that:
\begin{itemize}\item $l_a$ and $i_{a_1\ldots a_n}$ act by infinitesimal canonical transformations,
          generated by some BV Hamiltonians $\underline{l_a}$ and $\underline{i_{a_1\ldots a_n}}$\item The differential $d_{D\mathfrak{h}}$ acts as a second order differential operator, with the leading symbol
            defined by the odd Poisson bivector (the leading symbol of the odd Laplace operator)\end{itemize}

Eq. (\ref{EntersI}) becomes:
\begin{align} r_{W_{\mathfrak h}\otimes {\mathfrak a}}\langle d_{C\mathfrak h}\rangle \;=\;
 & d_{W_{\mathfrak h}} + r_{\mathfrak a}\langle d_{D\mathfrak h}\rangle +
             {\bf\theta}^a \{\underline{l_a},\_\} - \sum {\bf t}^{a_1}\cdots {\bf t}^{a_n}
             \{\underline{i_{a_1\ldots a_n}},\_\}
\label{ActionOfDCgOnWa} \\ \mbox{where \hspace{2.00000ex}}
 & r_{\mathfrak a}\langle d_{D\mathfrak h}\rangle = \Delta^{(0)} + \{S_{\rm BV},\_\}
\label{DifferentialOfDh} \end{align}
The action of $\iota\langle x\rangle$ for $x\in {\mathfrak h}$ is:
\begin{align} r_{W_{\mathfrak h}\otimes {\mathfrak a}}\langle\iota\langle x\rangle\rangle \;=\;
 & x^a{\partial\over\partial {\bf\theta}^a}
\label{ActionOfIotaOnWa} \end{align}
Every representation of $C\mathfrak h$ defines a representation of $D\mathfrak h$,
where the action of $i_{a_1a_2\ldots}$ is zero. 
In this case Eqs (\ref{ActionOfDCgOnWa}) and (\ref{ActionOfIotaOnWa})
may be called ``Cartan model of equivariant cohomology of $\mathfrak a$''.
But for our construction, we do not need the representation of $C{\mathfrak h}$;
we only need a representation of $D{\mathfrak h}$, with the differential $d_{D\mathfrak h}$ defined by Eq. (\ref{DifferentialOfDh}).
This already allows us to
define (following \cite{Alekseev:2010gr}) the action of $C{\mathfrak h}$ on $W_{\mathfrak h}\otimes {\mathfrak a}$. 

The ``exponentiation method'', Eq. (\ref{ExpD}), constructs the equivariant half-density,
as we will now explain.

\subsection{Action of $C\mathfrak{h}$ on half-densities}\label{sec:ChOnHalfDensities}

As we have just explained, we might not be able to represent $C\mathfrak{h}$ by BV Hamiltonians;
we only require a representation of $D\mathfrak h$.
But we always have the action of $C\mathfrak h$ on half-densities $\mbox{Dens}_{1/2}(M)$.
Indeed, ${\cal L}\langle x\rangle$ acts as the Lie derivative along $l_a x^a$,
and $\iota\langle x \rangle$ acts as the multiplication by $\underline{l_a}x^a$.
The differential of $C\mathfrak h$ acts as $\Delta_{\rm can}$.
Therefore we can consider Weil (or Cartan) model of $\mbox{Dens}_{1/2}(M)$.

\subsection{Equivariant half-density}\label{sec:EquivariantHalfDensity}

It turns out that the exponentiation method (see Eq. (\ref{ExpD})) brings 
$r_{W_{\mathfrak h}\otimes {\mathfrak a}}\langle d_{C\mathfrak h}\rangle$
to a basic cocycle of the Weil model $W_{\mathfrak h}\otimes \mbox{Dens}_{1/2}(M)$. As in Eq. (\ref{ExpD}),
the \emph{equivariant Master Density} $\rho_{1/2}(A,F)$
corresponding to $r_{W_{\mathfrak h}\otimes {\mathfrak a}}\langle d_{C\mathfrak h}\rangle$
is defined as the
unique half-density $\rho_{1/2}({\bf\theta}, {\bf t})$ such that:
\begin{equation}
   r_{\mathfrak a}(d_{D\mathfrak h}) +
   {\bf\theta}^a \{\underline{l_a},\_\} - \sum {\bf t}^{a_1}\cdots {\bf t}^{a_n} \{\underline{i_{a_1\ldots a_n}},\_\}
   \;=\;
   \Delta_{\rho_{1/2}({\bf\theta},{\bf t})}
   \end{equation}
In Darboux coordinates, the correspondence $d\leftrightarrow \rho_{1/2}$ is,
literally, exponentiation:
\begin{equation}
     \rho_{1/2}({\bf \theta},{\bf t}) \;=\;
     \exp\left(
               S_{\rm BV} - {\bf \theta}^a \underline{l_a} + \sum {\bf t}^{a_1}\cdots {\bf t}^{a_n} \underline{i_{a_1\ldots a_n}}
               \right)
     \label{EquivariantRhoInDarboux}\end{equation}
It satisfies:
\begin{equation}
     (d_{W_{\mathfrak g}} + \Delta_{\rm can})\rho_{1/2}({\bf \theta},{\bf t}) = 0
     \label{QMEWeil}\end{equation}
Moreover, $\rho_{1/2}({\bf \theta},{\bf t})$ is basic in the Weil model:
\begin{equation}
     \left({\partial\over\partial{\bf \theta}^a} + \underline{l_a}\right)\rho_{1/2}({\bf \theta}, {\bf t}) = 0
     \label{HorizonthalWeil}\end{equation}
(remember that $C\mathfrak h$ acts on half-densities,
          $\iota\langle x\rangle$ acts as multiplication by $\underline{x}$,
          in this case multiplication by $\underline{l_a}$).
Since $\rho_{1/2}({\bf\theta},{\bf t})$ is base, the Cartan model is obtained
by removing $e^{-l\langle{\bf\theta}\rangle}$ from $\rho_{1/2}({\bf\theta},{\bf t})$:
\begin{equation}
     \rho^{\tt C}_{1/2}({\bf t}) \;=\;
     \exp\left(
               S_{\rm BV} + \sum {\bf t}^{a_1}\cdots {\bf t}^{a_n} \underline{i_{a_1\ldots a_n}}
               \right)
     \label{EquivariantRhoCartan}\end{equation}
It satisfies:
\begin{equation}
   \Delta_{\rm can}\rho^{\tt C}_{1/2}({\bf t}) = \underline{l\langle {\bf t}\rangle} \rho^{\tt C}_{1/2}({\bf t})
   \end{equation}

\subsection{Derivation using Kalkman formulas}\label{sec:ComparisonToUsualKalkman}

We will now give another derivation of Eq. (\ref{EquivariantRhoInDarboux}) using standard techniques of equivariant
cohomology. We will first review the Kalkman map, and then consider two applications of it.
The first is the relation between Weil and Cartan models in equivariant cohomology theory, and
another is Eq. (\ref{EquivariantRhoInDarboux}). We then discuss how they are similar but different.

\subsubsection{General Kalkman formula}\label{sec:GeneralKalkman}

Let us consider a differential graded
Lie superalgebra $\mathfrak G$ and a $C{\mathfrak G}$-module $\mathfrak X$ with a compatible differential
$d_{\mathfrak{X}}$. 
The generators of $C\mathfrak{G}$ are $\iota\langle\gamma\rangle$ and ${\cal L}\langle\gamma\rangle$, where
$\gamma\in\mathfrak{G}$. The differential $d_{\mathfrak{X}}$ should be compatible in the following sense:
\begin{align} d_{\mathfrak{X}}{\cal L}\langle\gamma\rangle v \;=\;
 & {\cal L}\langle d_{\mathfrak{G}} \gamma\rangle v
                   + (-)^{\bar{\gamma}} {\cal L}\langle \gamma\rangle d_{\mathfrak{X}}v
\nonumber{} \\ d_{\mathfrak{X}}\iota\langle\gamma\rangle v \;=\;
 & \iota\langle d_{\mathfrak{G}} \gamma\rangle v
                        + (-)^{\bar{\gamma}}{\cal L}\langle\gamma\rangle v
                        + (-)^{\bar{\gamma}+1}\iota\langle \gamma\rangle d_{\mathfrak{X}}v
\nonumber{} \end{align}
Then, for any odd ${\cal C}\in \mathfrak G$:
\begin{equation}
   e^{\iota\langle{\cal C}\rangle}d_{\mathfrak X} e^{-\iota\langle{\cal C}\rangle}
   =
   d_{\mathfrak X} + {\cal L}\langle {\cal C} \rangle -
   \iota\left\langle d_{\mathfrak G}{\cal C} + {1\over 2}[{\cal C},{\cal C}]\right\rangle
   \label{ExpIotaC}\end{equation}
In particular, if $\mathcal C$ satisfies the MC equation:
\begin{equation}
     d_{\mathfrak G}{\cal C} + {1\over 2}[{\cal C},{\cal C}] = 0
     \label{MCE}\end{equation}
then $e^{\iota\langle {\cal C}\rangle}$
intertwines $d_{\mathfrak X}$ with $d_{\mathfrak X} + {\cal L}\langle {\cal C} \rangle$.

We will now compare two different applications of Eq. (\ref{ExpIotaC}).
The first application is the relation between Cartan and Weil models of equivariant cohomology,
and the second is the construction of equivariant half-density. 

\subsubsection{Cartan and Weil models of equivariant cohomology}\label{sec:CartanAndWeil}

Let us consider:
\begin{align} {\mathfrak G} \;=\;
 & W_{\mathfrak h}\otimes C\mathfrak{h}
\label{GEq} \\ d_{\mathfrak G} \;=\;
 & d_{W_{\mathfrak h}} + d_{C\mathfrak{h}}
\label{DGEq} \\ {\mathfrak X} \;=\;
 & W_{\mathfrak h} \otimes (\mbox{differential forms})
\label{XEq} \\ d_{\mathfrak X} \;=\;
 & d_{W_{\mathfrak h}} + d_{\rm deRham}
\label{DXEq} \end{align}
The representation of $\mathfrak G$ on $\mathfrak X$ is defined using
the projection $CC\mathfrak{h} \rightarrow C\mathfrak{h}$.
This projection is defined as follows. For any Lie superalgebra $\mathfrak g$, we can think of elements of
$C\mathfrak g$ as maps $\mathbb{R}^{0|1} \rightarrow \mathfrak{g}$, \textit{i.e.} functions $x(\zeta)$ where $\zeta$ is a Grassmann odd parameter.
The double cone $CC\mathfrak{h}$ is the space of functions of two Grassmann parameters $\zeta_1$ and $\zeta_2$.
More precisely, elements of $C\mathfrak h$ are functions of $x(\zeta_2)$, and elements of $CC\mathfrak h$
are functions of $x(\zeta_1,\zeta_2)$. The projection  $CC\mathfrak{h} \rightarrow C\mathfrak{h}$ is defined
as a pullback of the diagonal map $\mathbb{R}^{0|1} \times \mathbb{R}^{0|1} \rightarrow \mathbb{R}^{0|1}$.
In other words, given $x(\zeta_1,\zeta_2)$ we put $\zeta_1 = \zeta_2 = \zeta$.
We use the following notations for the elements of $CC\mathfrak{h}$:
\begin{align} {\cal L}\langle l \langle x\rangle\rangle\;=\;
 & (\zeta_1,\zeta_2) \mapsto x
\nonumber{} \\ \iota\langle l\langle x\rangle \rangle \;=\;
 & (\zeta_1,\zeta_2) \mapsto \zeta_1 x
\nonumber{} \\ {\cal L}\langle i\langle x\rangle \rangle \;=\;
 & (\zeta_1,\zeta_2) \mapsto \zeta_2 x
\nonumber{} \\ \iota\langle i\langle x\rangle\rangle \;=\;
 & (\zeta_1,\zeta_2) \mapsto \zeta_1\zeta_2 x
\nonumber{} \end{align}
The map $CC\mathfrak{h} \rightarrow C\mathfrak{h}$ commutes with the differential.
Indeed, the differential of $C\mathfrak h$ is $\partial\over\partial\zeta$, and the
differential of $CC\mathfrak h$ is ${\partial\over\partial\zeta_1} + {\partial\over\partial\zeta_2}$.
Under the diagonal map, $\partial\over\partial\zeta$ becomes
${\partial\over\partial\zeta_1} + {\partial\over\partial\zeta_2}$.

The generators of $C\mathfrak{h}$ will be denoted $l\langle x\rangle$ and $i\langle x\rangle$.
We take ${\cal C} = l\langle{\bf\theta}\rangle + i\langle{\bf t}\rangle$.
The MC equation is satisfied:
\begin{equation}
   (d_{W_{\mathfrak h}} + d_{C\mathfrak{h}}){\cal C} + {1\over 2}[{\cal C},{\cal C}] = 0
   \label{MCforKalkman}\end{equation}
Eq. (\ref{ExpIotaC}) in this case involves the operators ${\cal L}\langle l\langle x\rangle\rangle$,
$\iota\langle l\langle x\rangle\rangle$, ${\cal L}\langle i\langle x \rangle\rangle$
and $\iota\langle i\langle x\rangle\rangle$ which define the representation of $CC\mathfrak{h}$.

\subsubsection{Equivariant BV}\label{sec:EquivariantBV}

In this case we take:
\begin{align} {\mathfrak G} \;=\;
 & W_{\mathfrak h}\otimes D\mathfrak{h}
\label{GBV} \\ d_{\mathfrak G} \;=\;
 & d_{W_{\mathfrak h}}
\label{DGBV} \\ {\mathfrak X}\;=\;
 & W_{\mathfrak h}\otimes\mbox{Dens}_{1/2}M
\label{XBV} \\ d_{\mathfrak X}\;=\;
 & d_{W_{\mathfrak h}} + \Delta_{\rm can}
\label{DXBV} \end{align}
We act by LHS and RHS of Eq. (\ref{ExpIotaC}) to a vector $\rho_{1/2}\in \mbox{Dens}_{1/2}M$ such that:
\begin{equation}
   {\cal L}\langle x\rangle \rho_{1/2} = \iota\langle d_{D\mathfrak h} x\rangle \rho_{1/2}
   \label{LxVsIDx}\end{equation}
Eq. (\ref{LxVsIDx}) was discussed in \cite{Mikhailov:2020lzo}. Then we require:
\begin{equation}
   (d_{D\mathfrak h} + d_{W_{\mathfrak h}}){\cal C} + {1\over 2}[{\cal C},{\cal C}] = 0
   \label{MCforBV}\end{equation}
The solution to Eq. (\ref{MCforBV}) is:
\begin{equation}
   {\cal C} = l\langle{\bf\theta}\rangle + i({\bf t})
   \end{equation}
Eq. (\ref{ExpIotaC}) gives:
\begin{equation}
   (\Delta_{\rm can} + d_{W_{\mathfrak h}})e^{-\iota\langle{\cal C}\rangle}\rho_{1/2} =
   e^{-\iota\langle{\cal C}\rangle}(\Delta_{\rm can} + d_{W_{\mathfrak h}})\rho_{1/2} = 0
   \end{equation}

\subsubsection{Comparison}\label{sec:Comparison}

Eq. (\ref{MCforBV}) looks very similar to Eq. (\ref{MCforKalkman}). But unlike Eq. (\ref{MCforKalkman}),
it is \emph{not} a particular case of Eq. (\ref{MCE}).
A direct application of Eq. (\ref{MCE}) would give
$d_{W_{\mathfrak{h}}}{\cal C} + {1\over 2}[{\cal C},{\cal C}] = 0$,
which is different from Eq. (\ref{MCforBV}).
There is an action of $W_{\mathfrak{h}}\otimes CD\mathfrak{h}$ on $W_{\mathfrak{h}}\otimes\mbox{Dens}_{1/2}M$,
but $d_{D\mathfrak{h}}$ is missing from Eq. (\ref{DGBV}). 
In Section \ref{sec:CartanAndWeil} the $d_{deRham}$ corresponds
to the action of ${\partial\over\partial\zeta_1} + {\partial\over\partial\zeta_2}$, \textit{i.e.}
the sum of ``inner'' and ``outer'' differentials of $CC\mathfrak{h}$.
While in Section \ref{sec:EquivariantBV} we use $CD\mathfrak{h}$, but
there is no projection to $C\mathfrak{h}$ or $D\mathfrak{h}$ involved, and $\Delta_{\rm can}$ only
induces the ``outer'' differential. The $d_{D\mathfrak{h}}$ is missing from Eq. (\ref{DGBV}).
Instead, we take the ${\cal L}\langle{\cal C}\rangle$ term from Eq. (\ref{ExpIotaC}), ``convert''
it into $d_{D\mathfrak{h}}{\cal C}$ using Eq. (\ref{LxVsIDx}), and include in the differential of the MC equation.
As a result, Eq. (\ref{MCforBV}) is similar to Eq. (\ref{MCforKalkman}).

\subsection{Deformation complex}\label{sec:EquivariantDeformationComplex}

Eq. (\ref{HorizonthalWeil}) might appear counter-intuitive,
as $r_{W_{\mathfrak h}\otimes {\mathfrak a}}\langle d_{C\mathfrak h}\rangle$
(defined in Eq. (\ref{ActionOfDCgOnWa}))
corresponds to the Cartan model, and not Weil model.

Let us consider the deformation complex.
Infinitesimal deformations of $\rho_{1/2}({\bf \theta}, {\bf t})$ can be 
written as $V({\bf \theta}, {\bf t})\rho_{1/2}({\bf \theta}, {\bf t})$ where $V({\bf \theta}, {\bf t})$
is a function on $M$. Then:
\begin{equation}
     \iota_{W_{\mathfrak h}\otimes \mathfrak{a}}\langle l_a\rangle (V\rho_{1/2}) =
     \left({\partial\over\partial{\bf \theta}^a} + \underline{l_a}\right)(V\rho_{1/2}) =
     V\left({\partial\over\partial{\bf \theta}^a} + \underline{l_a}\right)\rho_{1/2} +
     \left({\partial\over\partial{\bf \theta}^a}V\right)\rho_{1/2}
     \label{IotaOnDeformedRho}\end{equation}
Therefore $\iota_{W_{\mathfrak h}\otimes \mathfrak{a}}\langle l_a\rangle$ acts on deformations as in the Cartan model:
\begin{equation}
     \iota_{W_{\mathfrak h}\otimes \mathfrak{a}}\langle l_a\rangle V = {\partial\over\partial{\bf \theta}^a}V
     \label{IotaOnDeformation}\end{equation}
In this sense, the map:
\begin{equation}
   V \mapsto V\rho_{1/2}
   \end{equation}
is a vartiation on a theme of ``Kalkman map'', the action by $e^{\iota\langle{\bf\theta}\rangle}$ relating
Weil and Cartan models. On the left hand side (Cartan), $V\in W_{\mathfrak h}\otimes\Pi\mbox{Fun}(M)$
is acted upon by the ``complicated'' differential
$r_{W_{\mathfrak h}\otimes {\mathfrak a}}\langle d_{C\mathfrak h}\rangle$, see Eq. (\ref{ActionOfDCgOnWa}),
and $\iota$ acts as $\partial\over\partial{\bf\theta}$.
On the right hand side (Weil), $V\rho_{1/2}\in W_{\mathfrak h}\otimes \mbox{Dens}_{1/2}(M)$
is acted upon by the ``canonical'' differential
$\Delta_{\rm can} + d_W$, and $\iota$ is $\underline{l} + {\partial\over\partial{\bf\theta}}$. 
Both describe deformations of the equivariant half-density $\rho_{1/2}$.
This is somewhat similar to the map $\Phi$ constructed in \cite{Alekseev:2010gr}, but the construction here is simpler.
The difference is that our map $V\mapsto V\rho_{1/2}$ acts between different representations.
It maps functions to half-densities.

The $C\mathfrak h$-invariance conditions (for the deformation to be base) are:
\begin{align}  
 & {\partial\over\partial{\bf \theta}^a}V({\bf \theta},{\bf t}) = 0
\label{HorizonthalV} \\  
 & \left.{d\over ds}\right|_{s=0} V\left({\rm Ad}(e^{s\xi}){\bf t}\right) = \{\underline{\xi}, V({\bf t})\}
\nonumber{} \end{align}

\section{Integration over families of Lagrangian submanifolds}\label{IntegrationOverFamilies}

Let us fix a Lagrangian submanifold $L$, and consider all Lagrangian
submanifolds which can be obtained from $L$ by canonical transformations.
This is an ``infinite-dimensional supermanifold''.
We will call it $\rm LAG$:
\begin{align} {\rm LAG}\;=\;
 & \{gL|g\in G\}
\nonumber{} \end{align}
Here $G$ is the group of canonical transformations of $M$.

Consider the following pseudo-differential form (PDF) on $\rm LAG$:
\begin{equation}
   \Omega = \int_{gL}\exp\left(\underline{dgg^{-1}} + S_{\rm BV}\right)
   \label{Omega}\end{equation}
This is a closed form. This can be seen by rewriting it equivalently as follows:
\begin{equation}
   \Omega = \int_L\exp\left(S_{\rm BV}\circ g + \underline{g^{-1}dg}\right)
   \end{equation}
The $\mathfrak{h}$-equivariant analogue of $\Omega$ is:
\begin{align} \Omega({\bf \theta},{\bf t}) \;=\;
 & \int_{gL}\exp\left(
                           S_{\rm BV}
                           +
                           \underline{dgg^{-1} - l\langle{\bf \theta}\rangle + i({\bf t})}
                           \right)\;=
\label{EquivariantOmega} \\ \;=\;
 & \int_L\exp\left(S_{\rm BV}\circ g + \underline{g^{-1}dg} - \underline{(l\langle{\bf \theta}\rangle - i({\bf t}))}\circ g\right)
\nonumber{} \end{align}
The construction of $\Omega$, as we now presented it, has the following defect.
We want to define a PDF on $\rm LAG$, not on $G$. Therefore $g$ is
only defined up to $g\sim gg_0$ where $g_0L = L$.
Therefore the restriction of $\underline{g^{-1}dg}$ on $L$ is only
defined up to a constant. This can be remedied in a number of ways, as
described in \cite{Mikhailov:2016myt},\cite{Mikhailov:2016rkp}. Here we will
assume that there is a ghost number symmetry, and consider only those
Lagrangian submanifolds which are invariant under the ghost number symmetry.
Then $\underline{g^{-1}dg}$ has ghost number $-1$, and cannot be constant.

We will use $g$ to parametrize $\rm LAG$, keeping an eye on $g\sim gg_0$.

\section{A BV interpretation of the integration of a differential form}\label{IntegrationOfForm}

\subsection{Some lagrangian submanifolds in $\Pi T^*\Pi TY$}\label{sec:LPiTPiT}

Let us consider a PDF $\alpha$ on a supermanifold $Y$. Let $\underline{d}\in \mbox{Fun}(\Pi T^*(\Pi TY))$ be the generating
function of the de Rham differential $d$ (considered as an odd vector field on $\Pi TY$).

For an oriented submanifold $X\subset Y$, we can define
$\int_X\alpha$ as $\int_{L_X} e^{\underline{d}}\alpha$ where $L_X\subset \Pi T^*(\Pi TY)$ is the conormal bundle
to $\Pi TY|_X \subset \Pi TY$. In local coordinates, $L_X$ can be described as follows.
We let $y\in Y$ and its antifield $y^{\star}$ run in the conormal bundle
of $X\subset Y$. At the same time, $dy$ runs free and $(dy)^{\star} = 0$.

To summarize:
\begin{equation}
   \int_X\alpha = \int_{L_X} e^{\underline{d}}\alpha
   \label{IntegralOfFormInBV}\end{equation}

\subsection{Some canonical transformations}\label{sec:SomeCanonicalTransformations} 

Let us consider, in this context, canonical transformations. For any  $\beta \in \mbox{Fun}(\Pi TY)$,
we can consider $\beta$ to be a function on $\Pi T^* \Pi TY$ (as the pullback under the projection) and compute
the Lie derivative of the half-density $e^{\underline{d}}\,\alpha$ under its Hamiltonian flow:
\begin{equation}
   {\cal L}_{\{\beta,\_\}} \left(e^{\underline{d}}\,\alpha\right) \,=\, (-)^{\bar{\beta}}e^{\underline{d}}\,d\beta\;\alpha
   \end{equation}
(This flow deforms $L_X$ away from $(dy)^{\star}=0$, showing that there is nothing we can get from this construction,
      beyond an integral of a differential form.)
For a vector field $v\in\mbox{Vect}(Y)$, let ${\cal L}_v$ be the Lie derivative (a vector field on $\Pi TY$),
and $\underline{{\cal L}_v}$ its BV Hamiltonian. The Lie derivative of the half-density along $\{\underline{{\cal L}_v},\_\}$ is:
\begin{equation}
   {\cal L}_{\{\underline{{\cal L}_v},\_\}} \left(e^{\underline{d}}\alpha\right)
   =
   - \underline{{\cal L}_v}\Delta_{\rm can}\left(e^{\underline{d}}\alpha\right)
   + \Delta_{\rm can}\left(\underline{{\cal L}_v} e^{\underline{d}}\alpha\right)
   =
   e^{\underline{d}}{\cal L}_v\alpha
   \end{equation}

\subsection{Universal canonical transformation}\label{sec:UniversalCanonicalTransformation}

Let us consider the extended BV phase space:
\begin{equation}
   \widehat{M} = M\times \Pi T^* \Pi T G
   \end{equation}
where $G$ is the group of canonical transformations of $M$ (or perhaps a subgroup of it).
We define the universal canonical transformation $\Gamma$ as follows:
\begin{equation}
     (m,\; g,\; dg,\; g^{\star}, (dg)^{\star})
     \stackrel{\Gamma}{\mapsto}
     (gm,\; g,\;  dg,\; g^{\star} + \alpha,\;  (dg)^{\star})
     \label{UniversalCanonicalTransformation}\end{equation}
where $\alpha$ is such that:
\begin{equation}
     dg(g^{\star} + \alpha) = dgg^{\star} + \underline{g^{-1}dg}
     \label{DefAlpha}\end{equation}
(Remember that $\underline{\xi}$ denotes the BV Hamiltonian of a vector field $\xi$.
          The left-invariant form $g^{-1}dg$ is a map $\Pi T G \rightarrow \mbox{Vect}(M)$,
          therefore $\underline{g^{-1}dg}$ is a function on $\Pi TG \times M$.)
          
The $\Omega$ of Eq. (\ref{Omega}) can be interpreted as a half denisty on $\widehat{M}$
\begin{equation}
     \exp\left(S_{\rm BV} + \underline{d\;}_{\Pi T^*\Pi TG}\right)\circ \Gamma
     \label{OmegaVsGamma}\end{equation}

\section{Restriction of gauge symmetry to a subgroup}\label{RestrictionToH0}

\subsection{General story}\label{sec:GeneralStory}

Consider a manifold $X$ with an action of the group $H$, and a subgroup $H_0\subset H$.
The Weil algebra of $\mathfrak{h}_0$ is the factoralgebra of the Weil algebra of $\mathfrak{h}$ by a differential ideal:
\begin{equation}
   W_{\mathfrak{h}_0} = W_{\mathfrak{h}}/{\cal I}
   \end{equation}
Let $\Omega^{\mathfrak{h}}({\bf\theta},{\bf t})$ be an $\mathfrak{h}$-equivariantly closed form on $X$, and $\Omega^{\mathfrak{h}_0}({\bf\theta}_0,{\bf t}_0)$ its restriction on $\mathfrak{h}_0\subset\mathfrak{h}$:
\begin{equation}
   \Omega^{\mathfrak{h}_0} = \Omega^{\mathfrak{h}} \;\mbox{mod}\; {\cal I}\otimes \Omega^{\bullet}(X)
   \end{equation}
In terms of connections and base forms, suppose that:
\begin{itemize}\item $a_0$ is a connection on $X\rightarrow H_0\backslash X$, and $\Omega_{{\tt base}[a_0]}^{\mathfrak{h}_0}$ the corresponding $\mathfrak{h}_0$-base form\item $a$ is a connection on $X\rightarrow H\backslash X$, with
          $\Omega_{{\tt base}[a]}^{\mathfrak{h}}$ the corresponding $\mathfrak{h}$-base form\end{itemize}
Then:
\begin{equation}
     \Omega_{{\tt base}[a]}^{\mathfrak{h}} = \Omega_{{\tt base}[a_0]}^{\mathfrak{h}_0} + d(\ldots)
     \label{GeneralRestorationOfH}\end{equation}
To see this, let us introduce a parameter $\tau$ and the following family of $\mathfrak{h}_0$-connections on
$X\times \mathbb{R}_{\tau}$ and their curvatures:
\begin{align}  
 & \hat{a} = a_0 + \tau(a - a_0)
\nonumber{} \\  
 & \hat{f} = d_{X\times \mathbb{R}_{\tau}}\hat{a} + {1\over 2} [\hat{a},\hat{a}]
\nonumber{} \end{align}
We observe that $\hat{a}$ and $\hat{f}$ satisfy the Weil algebra and we can substitute them for
$\bf\theta$ and $\bf t$, obtaining a closed form:
\begin{equation}
   d_{X\times\mathbb{R}_{\tau}} \Omega^{\mathfrak{h}}(\hat{a},\hat{f}) = 0
   \end{equation}
In particular:
\begin{equation}
   {\partial\over\partial\tau}\Omega^{\mathfrak{h}}(\hat{a},\hat{f}) =
   d\left({\partial\over\partial d\tau}\Omega^{\mathfrak{h}}(\hat{a},\hat{f})\right)
   \end{equation}
This implies Eq. (\ref{GeneralRestorationOfH}).

We will now develop this idea in the BV context.

\subsection{Varying the gauge group in BV formalism}\label{sec:VaryingGaugeGroupBV}

Let $a_0$ be a connection in the $H_0$-principal bundle
${\rm LAG}\longrightarrow H_0\backslash {\rm LAG}$:
\begin{align}  
 & a_0\in\Omega^1({\rm LAG},\mathfrak{h}_0)
\label{TypeOfA0} \\  
 & \iota_{v_{\xi}}a_0 = \xi\quad \forall \xi \in \mathfrak{h}_0
\nonumber{} \\  
 & {\cal L}_{v_{\xi}}a_0 = [\xi,a_0]\quad \forall\xi\in\mathfrak{h}_0
\nonumber{} \end{align}
We will start by using the trick of
Section \ref{IntegrationOfForm} to replace $\Omega$ with the half-density
on $(\Pi T^*\Pi TG)\times M$:
\begin{equation}
   \rho = \exp\left(
                    S_{\rm BV} - \underline{l\langle a_0\rangle} + \underline{i(f_0)} +
                    \underline{d_G}
                    \right)\circ\Gamma
   \label{BigRho}\end{equation}
This defines $\rho$ as a half-density on $(\Pi T^*\Pi T{\rm LAG})\times M$.
But ${\rm LAG}$ does not have closed cycles. We use a closed cycle in $H_0\backslash {\rm LAG}$, not in $\rm LAG$.
Therefore, we need a half-density on  $(\Pi T^*\Pi T(H_0\backslash {\rm LAG}))\times M$. Indeed, $\rho$ can be considered
as a half-density on  $(\Pi T^*\Pi T(H_0\backslash {\rm LAG}))\times M$, by the BV Hamiltonian reduction. We first put to zero
the BV Hamiltonians generating the action of $\Pi TH_0$ (a constraint on $g^{\star}$ and $[dg]^{\star}$), and then factor out
by this action. This is possible because $\rho$ is  $H_0$-base as a PDF on $\rm LAG$.

Let us introduce an extra parameter $\tau\in \mathbb{R}$ and denote:
\begin{align} a'(\tau)\;=\;
 & a_0 + \tau(a - a_0)
\nonumber{} \\ f'(\tau,d\tau)\;=\;
 & da' + {1\over 2}[a',a'\,]
\nonumber{} \end{align}
where the differential $d$ in $da'$ includes $d\tau{\partial\over\partial\tau}$.
We will now use the following general construction.
Consider an odd symplectic manifold $X$ and add a pair of Darboux conjugate 
coordinates $\tau$ and $\tau^{\star}$. We get the new odd symplectic manifold:
\begin{equation}
   X\times \Pi T^* \mathbb{R}_{\tau}
   \end{equation}
For any function ${\cal F}\in \mbox{Fun}(X\times \mathbb{R}_{\tau})$, there is the  canonical transformation:
\begin{align}  
 & \Phi\;:\;X\times \Pi T^* \mathbb{R}_{\tau}\longrightarrow X\times \Pi T^* \mathbb{R}_{\tau}
\nonumber{} \\  
 & \Phi
        \left[\begin{array}{c}
                    \tau \cr \tau^{\star} \cr x
                    \end{array}\right]
        =
        \left[\begin{array}{c}
                    \tau \cr
                    \tau^{\star} + {\cal F}(\tau,x) \cr
                    \gamma_{\tau}(x)
                    \end{array}\right]
\label{BigPhi} \end{align}
where $\gamma_{\tau}$ is a family of canonical transformations of $M$, parameterized by $\tau$,
defined from the equations:
\begin{align} \underline{\gamma^{-1}_{\tau}{\partial\over\partial\tau}\gamma_{\tau}} \;=\;
 & {\cal F}(\tau,\_)
\label{GammaDot} \\ \gamma_{\tau = 0} \;=\;
 & \mbox{id}
\label{InitialConditionOnGamma} \end{align}
More explicitly:
\begin{equation}
     \gamma_{\tau} = P\exp\left(\int_0^{\tau} d\tilde{\tau} \{{\cal F(\tilde{\tau},\_)},\_\}\right)
     \label{ExplicitGamma}\end{equation}
This construction is a special case of
Section \ref{sec:UniversalCanonicalTransformation},
namely the restriction from $G$ to a curve $\{\gamma_{\tau}|\tau\in {\bf R}\}\subset G$.

In particular, consider $X = (\Pi T^* \Pi TG)\times M$ and:
\begin{equation}
   {\cal F} = {\partial\over\partial d\tau}\underline{i(f')}
   \end{equation}
Consider the following ``extended'' half-density
on $(\Pi T^* \Pi T(G\times\mathbb{R}_{\tau}))\times M$:
\begin{equation}
   \widehat{\rho} = \exp\left(S_{\rm BV} -
                                \underline{l\langle a'\rangle} + \underline{i(f')} +
                                \underline{d_{G\times \mathbb{R}_{\tau}}}
                                \right)\circ\Gamma
   \end{equation}
It satisfies the QME on $\Pi T^* \Pi T(G\times\mathbb{R}_{\tau})\times M$. Since $\Phi$ is a canonical
transformation, the following half-density
also satisfies the QME:
\begin{equation}
   \widetilde{\rho} = \exp\left(S_{\rm BV} -
                                  \underline{l\langle a'\rangle} + \underline{i(f')} +
                                  \underline{d_{G\times \mathbb{R}_{\tau}}}
                                  \right)\circ\Phi^{-1}\circ\Gamma
   \end{equation}
Moreover, $\widetilde{\rho}$, as $\rho$, does descend by the BV Hamiltonian reduction to a half-density
in $\Pi T^* \Pi T ((H_0\backslash G)\times \mathbb{R}_{\tau})\times M$.
This is because $\Phi$ is a well-defined canonical transformation of  $\Pi T^* \Pi T ((H_0\backslash G)\times \mathbb{R}_{\tau})\times M$. Indeed,  $a'(\tau)$ is an $\mathfrak{h}_0$-connection for all values of $\tau$, and therefore
${\cal F}\circ g$ is $\mathfrak{h}_0$-base as a PDF on $G$. Therefore it descends to a PDF on $H_0\backslash G$. Moreover, it can be considered
a PDF on $H_0\backslash {\rm LAG}$, since $a'$ and $f'$ are defined as forms on $\rm LAG$ (see Eq. (\ref{TypeOfA0})).
                                                                                        
We observe that:
\begin{equation}
   d\tau{\partial\over\partial d\tau}\underline{i(f')} + d\tau\tau^{\star} =
   (d\tau\tau^{\star})\circ \Phi
   \end{equation}
Therefore
$\exp\left(S_{\rm BV} -
               \underline{l\langle a'\rangle} + \underline{i(f')} +
               \underline{d_{{G}\times \mathbb{R}_{\tau}}}
               \right)\circ\Phi^{-1}\circ\Gamma$
contains $d\tau$ and $\tau^{\star}$ only through the multiplicative factor $e^{d\tau\tau^*}$. Then QME implies:
\begin{align}  
 & {\partial\over\partial\tau}
        \left[
              \exp\left(S_{\rm BV} -
                          \underline{l\langle a'\rangle} + \underline{i(f')} +
                          \underline{d_{G\times \mathbb{R}_{\tau}}}
                          \right)\circ\Phi^{-1}\circ\Gamma
              \right] = 0
\nonumber{} \end{align}
Restricting to $d\tau = 0$ and taking $\tau=0$ then $\tau=1$, this implies:
\begin{equation}
     \exp\left(S_{\rm BV} -
                 \underline{l\langle a\rangle} + \underline{i(f)} +
                 \underline{d_G}
                 \right)\circ\gamma_1^{-1}\circ\Gamma
     \;=\;
     \exp\left(S_{\rm BV} -
                 \underline{l\langle a_0\rangle} + \underline{i(f_0)} +
                 \underline{d_G}
                 \right)\circ\Gamma
     \label{FlowWithPhi}\end{equation}
On the other hand:
\begin{align}  
 & \exp\left(S_{\rm BV} -
                    \underline{l\langle a\rangle} + \underline{i(f)} +
                    \underline{d_G}
                    \right)
                 \circ{\partial\over\partial\tau}\gamma_{\tau}^{-1}\circ\Gamma
                 \;=
\label{InsertionOfDerivativeOfPhi} \\ \;=\;
 & \Delta_{\rm can}
                  \left[
                        \left(
                              \exp\left(S_{\rm BV} -
                                          \underline{l\langle a\rangle} + \underline{i(f)} +
                                          \underline{d_G}
                                          \right)
                              \underline{{\partial\over\partial d\tau}i(f')}
                              \right)
                        \circ\gamma_{\tau}^{-1}\circ\Gamma\right]
\nonumber{} \end{align}
where $\Delta_{\rm can}$ is the canonical operator on $\Pi T^* \Pi T (H_0\backslash G)\times M$.
Therefore $\gamma_1^{-1}$ can be replaced with $\rm id$, in the following sense:
\begin{align}  
 & \exp\left(S_{\rm BV} -
                    \underline{l\langle a\rangle} + \underline{i(f)} +
                    \underline{d_G}
                    \right)\circ\gamma_1^{-1}\circ\Gamma
        \;=\;
\nonumber{} \\ =\;
 & \exp\left(S_{\rm BV} -
                       \underline{l\langle a\rangle} + \underline{i(f)} +
                       \underline{d_G}
                       \right)\circ\Gamma + \Delta_{\rm can}(\ldots)
\label{Phi0Removed} \end{align}

\subsection{Unintegrated and integrated vertices}\label{sec:UnintegratedAndIntegrated}

Suppose that $V\in\mbox{Fun}(M)$ satisfies the following conditions:
\begin{align}  
 & \Delta_{\rho_{1/2}(0,0)} V = 0
\nonumber{} \\  
 & \{l\langle{\bf t}_0\rangle,V\} = 0 \mbox{ for all } {\bf t}_0\in \mathfrak{h}_0
\nonumber{} \\  
 & \{i({\bf t}_0),V\} = 0 \mbox{ for all } {\bf t}_0\in \mathfrak{h}_0
\nonumber{} \end{align}
Then, the product:
\begin{equation}
   V\;\rho_{1/2}({\bf \theta}_0, {\bf t}_0)
   \end{equation}
satisfies the $\mathfrak{h}_0$-equivariant Master Equation.
This may be understood as the first derivative of the measure with the $S_{\rm BV}$ deformed
to $S_{\rm BV} + \epsilon V$ w.r.to $\epsilon$ at $\epsilon = 0$.

Eq. (\ref{FlowWithPhi}) implies:
\begin{align}  
 & \left[
              (V\circ\gamma_1)
              \exp\left(S_{\rm BV} -
                          \underline{l\langle a\rangle} + \underline{i(f)} +
                          \underline{d_G}
                          \right)
              \right]\circ\gamma_1^{-1}\circ\Gamma\;=\;
\label{IntegratedVsUnintegratedDensity} \\ \;=\;
 & \left[
                 V
                 \exp\left(S_{\rm BV} -
                             \underline{l\langle a_0\rangle} + \underline{i(f_0)} +
                             \underline{d_G}
                             \right)
                 \right]\circ\Gamma
\nonumber{} \end{align}
Similarly to Eq. (\ref{Phi0Removed}):
\begin{align}  
 & \left[
              (V\circ\gamma_1)
              \exp\left(S_{\rm BV} -
                          \underline{l\langle a\rangle} + \underline{i(f)} +
                          \underline{d_G}
                          \right)
              \right]
             \circ\gamma_1^{-1}\circ\Gamma
             \;=\;
\nonumber{} \\ =\;
 & \left[
                 (V\circ\gamma_1)
                 \exp\left(S_{\rm BV} -
                             \underline{l\langle a\rangle} + \underline{i(f)} +
                             \underline{d_G}
                             \right)
                 \right]\circ\Gamma + \Delta_{\rm can}(\ldots)
\label{Phi0RemovedWithV} \end{align}
Given an integration cycle $C\;\subset\;H_0\backslash {\rm LAG}$,
\begin{align}  
 & \int_{\mathbb{L}_C\times L}
        \left[
              (V\circ\gamma_1)
              \exp\left(S_{\rm BV} -
                          \underline{l\langle a\rangle} + \underline{i(f)} +
                          \underline{d_G}
                          \right)
              \right]\circ\Gamma\;=\;
\label{Integrated} \\ \;=\;
 & \int_{\mathbb{L}_C\times L}
                \left[
                      V
                      \exp\left(S_{\rm BV} -
                                  \underline{l\langle a_0\rangle} + \underline{i(f_0)} +
                                  \underline{d_G}
                                  \right)
                      \right]\circ\Gamma
\label{Unintegrated} \end{align}
where $\mathbb{L}_C\subset \Pi T^*\Pi T(H_0\backslash {\rm LAG})$ is the Lagrangian
submanifold corresponding to $C$ by the construction of
Section \ref{IntegrationOfForm}.
Going back to the language of differential forms:
\begin{align}  
 & \int_C\int_L
        (V\circ\phi\circ g)
        \exp\left(\underline{g^{-1}dg}\right)
        \;\;
        \exp\left(
                  S_{\rm BV} - \underline{l\langle a\rangle} + \underline{i(f)}
                  \right)
        \circ g\;=\;
\label{IntegratedPDF} \\ \;=\;
 & \int_C\int_L
           (V\circ g)
           \exp\left(\underline{g^{-1}dg}\right)
           \;\;
           \exp\left(S_{\rm BV} - \underline{l\langle a_0\rangle} + \underline{i(f_0)}\right)
           \circ g
\label{UnintegratedPDF} \\ \mbox{where\hspace{1.00000ex}}
 & \phi\in\mbox{Map}\left(\Pi T (H_0\backslash H),G\right)
\nonumber{} \\  
 & \phi = P\exp\int_0^1 d\tau {\partial\over\partial d\tau}i(f')
\nonumber{} \end{align}
The integration cycle $C$ contains the vertical direction $H_0\backslash H$.

In Eq. (\ref{UnintegratedPDF}),
$\exp\left(\underline{g^{-1}dg}\right)
            \;\;
            \exp\left(S_{\rm BV} - \underline{l\langle a_0\rangle} + \underline{i(f_0)}\right)
            \circ g$
is $\mathfrak{h}_0$-base,
and $V\circ g$ is the insertion of the \emph{unintegrated} vertex operator. 

In Eq. (\ref{IntegratedPDF}),
$\exp\left(\underline{g^{-1}dg}\right)
            \;\;
            \exp\left(
                      S_{\rm BV} - \underline{l\langle a\rangle} + \underline{i(f)}
                      \right) \circ g$
is $\mathfrak{h}$-base. All vertical differentials come from $\phi$ entering as $V\circ \phi\circ g$.
The ``vertical'' integration of $V\circ \phi$ corresponds to the insertion of the \emph{integrated}
vertex operator corresponding to $V$. (``Vertical integration'' means integration along $H_0\backslash H$.)

\section{Integrated Vertex}\label{IntegratedVertex}

\subsection{Local trivialization of ${\rm LAG}\longrightarrow H\backslash {\rm LAG}$}\label{sec:LocalTrivialization}

Locally we can choose a section:
\begin{equation}
   s\;:\;H\backslash {\rm LAG} \longrightarrow {\rm LAG}
   \end{equation}
Each $g\in H_0\backslash {\rm LAG}$ can be written as a product:
\begin{align} g\;=\;
 & hs(m)
\nonumber{} \\ h\in
 & H_0\backslash H
\nonumber{} \\ m\in
 & H\backslash {\rm LAG}
\nonumber{} \end{align}
With this local trivialization, the $\mathfrak{h}$-connection $a$ can be written as follows:
\begin{align} a\;=\;
 & dh h^{-1} + hAh^{-1}
\nonumber{} \\ \mbox{where\hspace{1.00000ex}}
 & A\in \Omega^1(H\backslash{\rm LAG})\otimes \mathfrak{h}
\nonumber{} \end{align}

\subsection{Unperturbed measure in terms of local trivialization}\label{sec:UnperturbedMeasure}

Eq. (\ref{EquivariantOmega}) becomes:
\begin{align}  
 & \int_L
        e^{\underline{s^{-1}ds}}
        \exp\left(
                  S_{\rm BV} -
                  \underline{l\langle A\rangle}   +
                  \underline{i\left(dA + {1\over 2}[A,A]\right)}
                  \right)\circ s
\nonumber{} \end{align}
Gauge invariance is manifested by the $h$-independence of this measure.

\subsection{$\gamma_1$ in local trivialization}\label{sec:Phi1Trivialization}

\begin{align} \gamma_1\circ g\;=\;
 & h\circ
        \left(
              P\exp\int_0^1
              d\tau
              {\partial\over\partial d\tau}
              i\Big(
                    d\tau (h^{-1}dh - h^{-1}a_0h + A)+\hat{F}
                    \Big)
              \right)\circ s(m)
\label{PhiInTrivialization} \\ \mbox{where\hspace{1.00000ex}}
 & \hat{F} = h^{-1}\Big((1-\tau)f_0 + \tau f + {1\over 2} \tau(1-\tau)[a-a_0,a-a_0]\Big)h\;=
\nonumber{} \\  
 & \phantom{\hat{F}} =
        (1-\tau)h^{-1}f_0h +
        \tau \left(dA + {1\over 2}[A,A]\right) \;+
\nonumber{} \\  
 & \phantom{\hat{F} =}
                + {1\over 2} \tau(1-\tau) [h^{-1}\nabla^{[a_0]}h + A,h^{-1}\nabla^{[a_0]}h + A]
\nonumber{} \end{align}

\subsection{Local integrated vertex}\label{sec:LocalIntegratedVertex}

Locally on $\rm LAG$ we can choose connection so that $A=0$. Then:
\begin{align}  
 & V\circ\gamma_1\circ h \;=\;
\nonumber{} \\ \;=\;
 & \left(
                 P\exp\int_0^1
                 d\tau
                 {\partial\over\partial d\tau}
                 i\Big(
                       d\tau h^{-1}\nabla^{[a_0]}h +
                       (1-\tau)h^{-1}f_0h +
                       {1\over 2}\tau(1-\tau)\left[h^{-1}\nabla^{[a_0]}h,h^{-1}\nabla^{[a_0]}h\right]
                       \Big)
                 \right)(V\circ h)
\label{IntegratedVertexInBVLanguage} \end{align}
This is a PDF in $H_0\backslash H$, and has to be integrated over a closed cycle in $H_0\backslash H$.
The result is the integrated vertex operator corresponding to $V$:
\begin{equation}
   U = \int_c V\circ\gamma_1
   \end{equation}
It satisfies the Master equation:
\begin{equation}
   \Delta_{\rm can}\left(Ue^{S_{\rm BV}}\right) = 0
   \end{equation}
We will now give a direct proof of this. The restriction of $\Omega_V^{\mathfrak{h}_0-{\tt base}}$ to
$H_0\backslash HL\;\subset \;H_0\backslash {\rm LAG}$ is:
\begin{equation}
   \Omega_V^{\mathfrak{h}_0-{\tt base}}|_{H_0\backslash HL}
   \;=\;
   \int_L
   (V\circ h)
   \exp\left(S_{\rm BV} + \underline{l\langle h^{-1}(dh h^{-1} - a_0)h\rangle  + i(h^{-1}f_0h)}\right)
   \end{equation}
We know by construction this is a closed form, \textit{i.e.} annihilated by $d_{H_0\backslash H}$.
Let us consider the half-density on $(\Pi T^*\Pi T (H_0\backslash H))\times M$ corresponding
to $\Omega_V^{\mathfrak{h}_0-{\tt base}}|_{H_0\backslash HL}$ by
the construction of
Section \ref{IntegrationOfForm}.
It is equal to:
\begin{align}  
 & (V\circ h)
        \exp\left(
                  S_{\rm BV} +
                  \underline{l\langle h^{-1}(dh h^{-1} - a_0)h\rangle  + i(h^{-1}f_0h)} +
                  \underline{d_{H_0\backslash H}}
                  \right)\;=
\nonumber{} \\ =\;
 & \left[
                 (V\circ\gamma_1)
                 \exp\left(
                           S_{\rm BV} +
                           \underline{d_{H_0\backslash H}}
                           \right)
                 \right]\circ\gamma_1^{-1}\circ \Gamma
\nonumber{} \end{align}
This satisfies the Master Equation by construction. As $\gamma_1$ is a canonical transformation,
and $\Gamma^{-1}\circ\gamma_1\circ\Gamma$ is a canonical transformation,
the following half-density on $(\Pi T^*\Pi T (H_0\backslash H))\times M$ also satisfies the Master Equation:
\begin{equation}
   \left[
         (V\circ\gamma_1)
         \exp\left(
                   S_{\rm BV} +
                   \underline{d_{H_0\backslash H}}
                   \right)
         \right]\circ\Gamma
   \end{equation}
Then pick a Lagrangian submanifold
$\mathbb{L}_c\in \Pi T^*\Pi T (H_0\backslash H)$ corresponding to a closed cycle $c\subset H_0\backslash H$
and integrate over it, as a half-density on $\Pi T^*\Pi T (H_0\backslash H)$.
This is essentially a BV push-forward:
\begin{equation}
   (\Pi T^*\Pi T (H_0\backslash H))\times M \longrightarrow M
   \end{equation}
The result satisfies the Master Equation on $M$. 
\footnote{
         In a typical application of BV formalism,  $\Pi T^*\Pi T(H_0\backslash H)$ would be considered
         ``fast degrees of freedom'', which we integrate out. 
         This gives an ``effective action'' on $M$ given by Eq. (\ref{IntegratedVertexInBVLanguage}).
         This terminology does not seem to be very appropriate in our context.
         Should one visualize the point of insertion running very fast all over the worldsheet?
         }

\section{Deformation of $i$}\label{DeformationOfI}

\subsection{Direct computation}\label{sec:DeltaIDirect}

Eq. (\ref{PhiInTrivialization}) contains the dependence on $F = dA + {1\over 2}[A,A]$.
After the integration over $h$, this computes the deformation of $i$ due to the insertion of $F$:
\begin{align} \delta_V i(F)\;=\;
 & \int_{H_0\backslash H}
                \Big(
                      P\exp\int_0^1
                      d\tau
                      {\partial\over\partial d\tau}
                      i\Big(
                            d\tau h^{-1}\nabla^{[a_0]}h \;+
                            %))
\nonumber{} \\  
 & \phantom{\int_{H_0\backslash H}
                      \Big(
                            P\exp\int_0^1
                            d\tau
                            {\partial\over\partial d\tau}
                            i\Big(
                            %))
                      }
                + (1-\tau)h^{-1}f_0h +
                {1\over 2}\tau(1-\tau)\left[h^{-1}\nabla^{[a_0]}h,h^{-1}\nabla^{[a_0]}h\right] \;+
\nonumber{} \\  
 & \phantom{\int_{H_0\backslash H}
                      \Big(
                            P\exp\int_0^1
                            d\tau
                            {\partial\over\partial d\tau}
                            i\Big(
                                  %))
                      }
                + F %((
                       \Big)
                      \Big)(V\circ h)
\label{IntegralForDeformationOfI} \end{align}

\subsection{Computation from the defining equation}\label{sec:DeltaIComputationFromEquation}

We have to find $\delta_V i({\bf t})$ such that:
\begin{equation}
   \{S_{\rm BV},\delta_V i({\bf t})\} + \{i({\bf t}),\delta_V i({\bf t})\}
   + \{U, i({\bf t})\} = 0
   \end{equation}
Here we will restrict ourselves with the case when $\mbox{dim}(H_0\backslash H) = 1$. The integration is one-dimensional,
let $u$ denote the coordinate on $H_0\backslash H$. Then the integrated vertex is:
\begin{equation}
   U = \int du \{ i_1\langle A_u \rangle , (V\circ f(u)) \}
   \end{equation}

Given a ${\bf t}\in \mathfrak{h}$
such that ${\bf t}\neq 0 \;\mbox{mod}\;\mathfrak{h}_0$,
let us define $\tilde{A}_u$ so that:
\begin{align}  
 & \tilde{A}_u - A_u \in \mathfrak{h}_0
\nonumber{} \\  
 & \tilde{A}_u \in {\mathbb R}{\bf t}
\nonumber{} \end{align}
We can expand $i({\bf t})$ in powers of $\bf t$:
\begin{equation}
   i({\bf t})
   = i_1\langle {\bf t}\rangle
   + i_2\langle {\bf t}\otimes {\bf t}\rangle
   + i_3\langle {\bf t}\otimes {\bf t}\otimes {\bf t} \rangle
   + \ldots
   \end{equation}
We claim that the variation of $i$ from the insertion of $U$ equals to:
\begin{equation}
   \delta i({\bf t})
   = \int du
   \left\{\;
   i_2\langle {\bf t}\bullet \tilde{A}_u \rangle
   + i_3\langle {\bf t}\bullet {\bf t}\bullet \tilde{A}_u\rangle
   + \ldots
   + i_n\langle {\bf t}^{\bullet (n-1)}\bullet \tilde{A}_u\rangle + \ldots
   \;,\;
   V\circ f
   \right\}
   \end{equation}
where $\bullet$ means symmetrized tensor product:
\begin{equation}
   v_1\bullet\cdots\bullet v_n =
   {1\over n!} \sum_{\sigma\in S_n} v_{\sigma_1}\otimes \cdots \otimes v_{\sigma_n}
   \end{equation}
This agrees with Eq. (\ref{PhiInTrivialization}). 

\subsection{Some observations}\label{sec:DeltaIObservations}

The case $\mbox{dim}(H_0\backslash H)=1$ has the following property.
If $i({\bf t})$ is a polynomial of degree $n$, it will remain a polynomial of degree $n$
after the deformation.
Moreover, the highest order term $i_n\langle {\bf t}^{\otimes n}\rangle$ does not receive correction.

Suppose that $i_n = 0$ for $n>1$, \textit{i.e.} $i({\bf t})$ is linear in $\bf t$.
Then $\delta_V i({\bf t})=0$, even when $\mbox{dim}(H_0\backslash H)\neq 1$.
This can be seen from Eq. (\ref{PhiInTrivialization}).
The curvatures would all come from the $F'$ in Eq. (\ref{PhiInTrivialization}),
but they only enter through $i_{n\geq 2}$.
At the same time, in this case  $\{U,i\langle{\bf t}\rangle\} = 0$, and therefore no need to correct $i({\bf t})$.
This observation might have an implication for the pure spinor superstring.
Although the BV formalism for the pure spinor superstring has not been fully developed
(only the $i_1$ has been constructed, in \cite{Mikhailov:2017mdo}, for the AdS background),
we do have some hints. The $i_1$ must be related to OPE with the $b$-ghost.
The pure spinor $b$-ghost (unlike the bosonic string $b$-ghost) does receive corrections
when we deform the background \cite{Chandia:2013ima}, \cite{Chandia:2019klv}. This seems to indicate that higher $i_n$ are  nonzero
in the pure spinor formalism.

\section*{Acknowledgments}

This work was supported in part by FAPESP thematic grants 2016/01343-7 and 2019/21281-4.

\def\cprime{$'$} \def\cprime{$'$}
\providecommand{\href}[2]{#2}\begingroup\raggedright\endgroup

\end{document}